\documentclass[prfluids,preprint]{revtex4-1}

\usepackage{graphicx}
\usepackage{psfrag}
\usepackage{amsmath}
\usepackage{amssymb}
\usepackage{wasysym}
\usepackage{color}
\usepackage{xcolor}
\usepackage{tikz}
\usepackage{tabularx}
\graphicspath{ {./Figures/} }

\begin{document}


\title{Evidence for multiplicative cascades as the origin of small-scale intermittency in turbulence}


\author{Alberto Vela-Mart\'in}
\email[]{alberto.vela.martin@zarm.uni-bremen.de}
\affiliation{Center of Applied Space Technology and Microgravity (ZARM), University of Bremen, 28359 Bremen}

\date{\today}

\begin{abstract}
This investigation presents evidence of the relation between the dynamics of intense events in small-scale turbulence and the energy cascade.
We use the generalised (H\"older) means to track the temporal evolution of intense events of the enstrophy and the dissipation
in direct numerical simulations of isotropic turbulence. We show that these events are modulated by large-scale fluctuations,
and that their evolution is consistent with a local multiplicative cascade, as hypothesised by a broad class of intermittency models of turbulence.
\end{abstract}




\maketitle

\section{Introduction}

Early experimental measurements of the velocity gradients in turbulent flows
revealed the `spottiness' of small-scale turbulence \cite{batchelor1949nature}, 
and anticipated a far more complex structure of the small scales than originally postulated in
the seminal work of Kolmogorov \cite{kolmogorov1941local}.
This complexity became evident by the large amount of data gathered since \citep{frisch1995turbulence}.
At variance with Kolmogorov's assumptions, the dissipation and the enstrophy (the square of the vorticity vector)
are known to concentrate in regions of the flow where they become orders of magnitude more intense
than the average, the more so the larger the Reynolds number, apparently without bound \citep{buaria2019extreme}. 
These intense events organise in structures of theoretical and practical relevance 
whose origin and dynamics are not well understood \citep{yeung2015extreme,buaria2020self}.


A persuasive explanation of this phenomenon, known as small-scale intermittency,
comes from models supported on the theory of the energy cascade \citep{frisch1995turbulence,biferale2003shell}.
A particularly intuitive and successful class of models are those based on multiplicative cascades \citep{kolmogorov1962refinement,mandelbrot1974intermittent,frisch1978simple},
which describe small-scale intense structures as generated by the successive uneven breakup of eddies through the cascade process.
These models seem well-suited to describe the geometry of the small scales, 
particularly those based on the multifractal formalism \cite{meneveau1987simple,meneveau1991multifractal}. 
But they have limitations.
They are essentially geometrical, lacking temporal dynamics, 
and their connection to the Navier--Stokes equations is unclear.
Most importantly, their success does not substantiate the phenomenological assumptions
on which they rest, which must be further tested against empirical evidence.



Despite extensive research, the relation between the energy cascade 
and the dynamics of intense events in the dissipative range is 
not fully supported by the available evidence.
Neither is this relation evident in the Navier--Stokes equations.
Quite the opposite, the evolution equation of the velocity gradients contains a local non-linear mechanism that naturally leads to their self-amplification \citep{vieillefosse1984internal,cantwell1992exact,li2005origin},
partially explaining the emergence of intense events without resorting to cascades.
Models based on a simplified representation of this mechanism predict---in some cases exceptionally well---the intermittent probability density function
of the velocity gradients \citep{kraichnan1990models,she1991physical,li2005origin,wilczek2009dynamical}. 
Moreover, the statistics of the velocity gradients are known to be non-Gaussian even for Reynolds numbers
in which a cascade in the Kolmogorov sense is difficult to conceive \citep{schumacher2007asymptotic}.
This evidence suggests that intense events may be generated and controlled by local mechanisms independent of the cascade,
as proposed for intense vortices \citep{she1991physical,jimenez1993structure,jimenez2000intermittency}.


In this work we present compelling empirical evidence 
that intense events in small-scale turbulence are 
a dynamical consequence of the energy cascade,
and that their evolution is consistent with a local multiplicative cascade, 
as implied by a wide class of intermittency models. 
This investigation is motivated by recent synchronisation experiments in the dissipative range of isotropic turbulence \citep{vela2021},
which point to the key role of inertial-scale dynamics in the evolution and formation of intense vortices. 
%

We elaborate on this idea by analysing the temporal fluctuations of the intense enstrophy and strain (dissipation) events
in direct numerical simulations. We relate these fluctuations to the fluctuations of the average dissipation, and of its large-scale surrogate, 
and interpret the results from a causal perspective, considering the strong large-to-small scale coupling revealed by the synchronisation experiments.
This work is supported on simulations at moderate Reynolds numbers, which, although smaller than the largest simulations currently available, 
span unprecedentedly long times, allowing access to unexplored physics of intense events in small-scale turbulence.

\section{Methods}

We study the temporal evolution of the enstrophy, $\Omega=\boldsymbol \omega^2$, 
and the square of the strain, $\Sigma=2\boldsymbol{S}:\boldsymbol{S}$, in an incompressible isotropic turbulent flow. 
Here $\boldsymbol \omega=\nabla\times\boldsymbol u$ and $\boldsymbol S=\tfrac{1}{2}(\nabla\boldsymbol u + \nabla\boldsymbol u^T)$ are the vorticity vector and the rate-of-strain tensor, respectively, and $\boldsymbol u$ is the velocity vector.
Note that $\Sigma$ is equivalent to the local energy dissipation rate.
To track the dynamics of the most intense events of $\Sigma$ and $\Omega$,
we use their generalised (H\"older) means with integer exponent $p$, hereafter $p$-means,
\begin{equation}
\Omega^{(p)}(t)=\langle \Omega^p\rangle^{1/p},\ \Sigma^{(p)}(t)=\langle\Sigma^p\rangle^{1/p},
\label{eq:1}
\end{equation}
where the brackets denote spatial averaging over the flow domain.
We remark that these quantities are instantaneous 
spatial averages that fluctuate in time. 
For $p>1$, the generalised means give more weight to the intense events of $\Sigma$ and $\Omega$, the more so the larger $p$,
as demonstrated by their inequality property, $\Omega^{(p-1)}\le \Omega^{(p)} \le \Omega^{(p+1)}$ (similar for $\Sigma$).
In particular, when $p\rightarrow\infty$, the $p$-means are equal to the maximum value of
$\Omega$ or $\Sigma$ within the flow domain.  
For $p=1$, we recover the space-averaged enstrophy and strain, 
and for negative $p$, the $p$-means capture the weak velocity gradients in the turbulent background, 
the more so the smaller $p$; when $p\rightarrow-\infty$, they are equal to the minimum value of $\Omega$ or $\Sigma$ within the domain.
To each $p$-mean, we assign the characteristic intensity of 
the structures that its represents. We define this intensity as
\begin{equation}
\omega^{(p)}=\frac{\overline{\langle \Omega^{p+1}\rangle}}{\overline{\langle \Omega^{p}\rangle}},\text{  }\sigma^{(p)}=\frac{\overline{\langle \Sigma^{p+1}\rangle}}{\overline{\langle \Sigma^{p}\rangle}},
\label{eq:2}
\end{equation}
where the bar denotes the average over time. 
These quantities represent the typical intensity of the events that contribute the most on average 
to $\Omega^{(p)}$ and $\Sigma^{(p)}$.
The characteristic time-scale of these events is given by
\begin{equation}
t_{\omega}^{(p)}=\frac{1}{\sqrt{\omega^{(p)}}},\text{   }t_{\sigma}^{(p)}=\frac{1}{\sqrt{\sigma^{(p)}}}.
\label{eq:22}
\end{equation}

The $p$-means are naturally related (at least in the case of $\Sigma$)
to the high-order moments of the dissipation field, which have been previously studied in the context of the 
scaling exponents in turbulence \citep{schumacher2007asymptotic,schumacher2014small}.
Our work extends these and similar investigations by including temporal analysis, 
which allows to probe not only the geometrical structure of the velocity gradients,
but also the dynamics that generates it.

%



\begin{table}
\begin{center}
\begin{tabularx}{0.8\textwidth}{c @{\extracolsep{\fill}}cccccccccc}
        $N$  &$Re_\lambda$ & $k_{max}\eta$ & $L/\eta$ & $T/t_{\eta}$ & $T_{s}/T$ & $\Delta t/t_\eta$ & $\omega^{(3)}$ & $\sigma^{(3)}$ & $\omega^{(5)}$ & $\sigma^{(5)}$ \\
        \hline
        128  &  72  &     2.0    &   39   &  9 & \bf{2130}   & 0.19 & 21 &  8 &  65 &  22  \\
        192  &  97  &     2.0    &   57   & 11 & \bf{2231}   & 0.25 & 29 & 10 &  98 &  30  \\
        256  & 120  &     2.0    &   75   & 14 & \bf{2276}   & 0.31 & 38 & 13 & 195 &  46  \\
        384  & 159  &     2.0    &  111   & 17 & \bf{2322}   & 0.41 & 53 & 17 & 239 &  66  \\
        512  & 195  &     2.0    &  148   & 21 & \bf{2341}   & 0.48 & 70 & 21 & 399 & 106  \\
\end{tabularx}
\end{center}
\caption{Main parameters of the simulations, where $N$ is the number of grid points in each direction,
$k_{\max}=\sqrt{2}/3N$ is the maximum Fourier wave-number magnitude resolved in the simulations,
$T_{s}$ is the total runtime of each simulation, and $\Delta t$ is the temporal resolution of the signals obtained from each simulation.
The characteristic intensities, $\omega^{(p)}$ and $\sigma^{(p)}$, are normalised with $\overline{\Omega^{(1)}}=\overline{\Sigma^{(1)}}$.
}
\label{fig:table}
\end{table}

We analyse the temporal evolution of $\Omega^{(p)}$ and $\Sigma^{(p)}$ for $-4<p<5$
in direct numerical simulations, which are now briefly introduced. 
We consider incompressible isotropic turbulence at different Reynolds numbers in the range of $Re_\lambda=\overline{U}\lambda/\nu=70$-$195$, 
where $\lambda=(15\nu/\overline{\varepsilon})^{1/2}\overline{U}$ is the Taylor microscale. 
Here $U(t)$ is the instantaneous root-mean-square of the velocity fluctuations calculated over the flow domain,
$\varepsilon(t)=\nu\Sigma^{(1)}(t)=\nu\Omega^{(1)}(t)$
is the instantaneous space-averaged energy dissipation, and $\nu$ is the kinematic viscosity.
The Kolmogorov length- and time-scales are $\eta=(\nu^3/\overline{\varepsilon})^{1/4}$ and $t_\eta=(\nu/\overline{\varepsilon})^{1/2}$.
The large-scale eddy-turnover time is $T=\overline{U}/L$, where $L$ is the integral length of the longitudinal velocity autocorrelation.
The flow is simulated in a triply periodic cubic box using a Fourier pseudo-spectral code, 
and linearly forced in the large scales to achieve a statistically steady state \cite{cardesa2017turbulent}. 
The volume of the computational domain is approximately $(5.2L)^3$.
Further details of the simulations are presented in table \ref{fig:table}. 
We use a standard spatial resolution in our simulations, $k_{\max}\eta=2$, 
where $k_{\max}$ is the largest resolved wavenumber in Fourier space,
and we have discarded resolution issues by comparing the $p$-means of simulations with $k_{\max}\eta=3$.
A particularity of the simulations in this work is that they span a very long time, approximately $2000T$ in all cases.
This is necessary to robustly characterise the evolution of the $p$-means.
We calculate the $p$-means on the fly with enough temporal resolution (see table \ref{fig:table}),
and with appropriate finite-precision arithmetic to avoid numerical problems.

The Reynolds numbers considered here show small-scale intermittency effects,
and, at least for $Re_\lambda>100$, an inertial range of scales \citep{cardesa2015temporal}.
In table \ref{fig:table},  we show the characteristic intensities of the events targeted by the $p$-means
for $p=3$ and $5$. 
They increase with $Re_\lambda$, and reach values up to 
$10^2$ times larger than the average enstrophy and strain. 

In the following, we compare the temporal evolution of $\Omega^{(p)}$ and $\Sigma^{(p)}$ with the temporal evolution of 
the average energy dissipation, $\varepsilon(t)$, and with the evolution of the instantaneous
surrogate energy dissipation \citep{Taylor35}, defined as
\begin{equation}
\varepsilon_s(t)=\frac{U(t)^3}{L(t)}
\end{equation}
where $L(t)$ is the integral scale of the velocity autocorrelation calculated at a particular instant in time.
Note that also $U(t)$ is a fluctuating quantity. Although the surrogate dissipation is a large-scale quantity,
it is related to the average dissipation, and $\overline{\varepsilon_s}=C_{s}\overline{\varepsilon}$, where $C_s$ is 
constant of order unity.


\section{Results}

\begin{figure}
\includegraphics[width=0.45\textwidth]{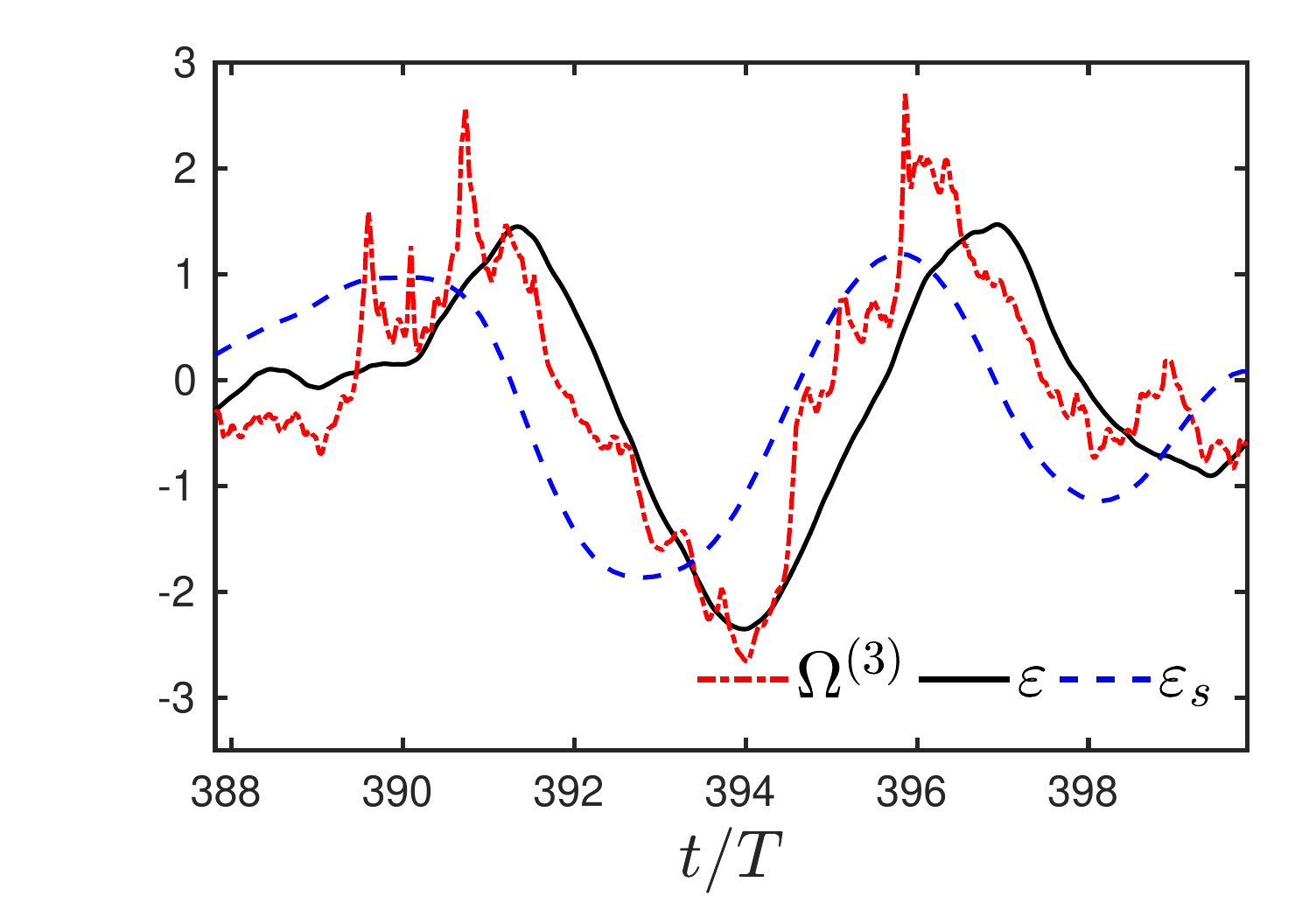}
\includegraphics[width=0.35\textwidth]{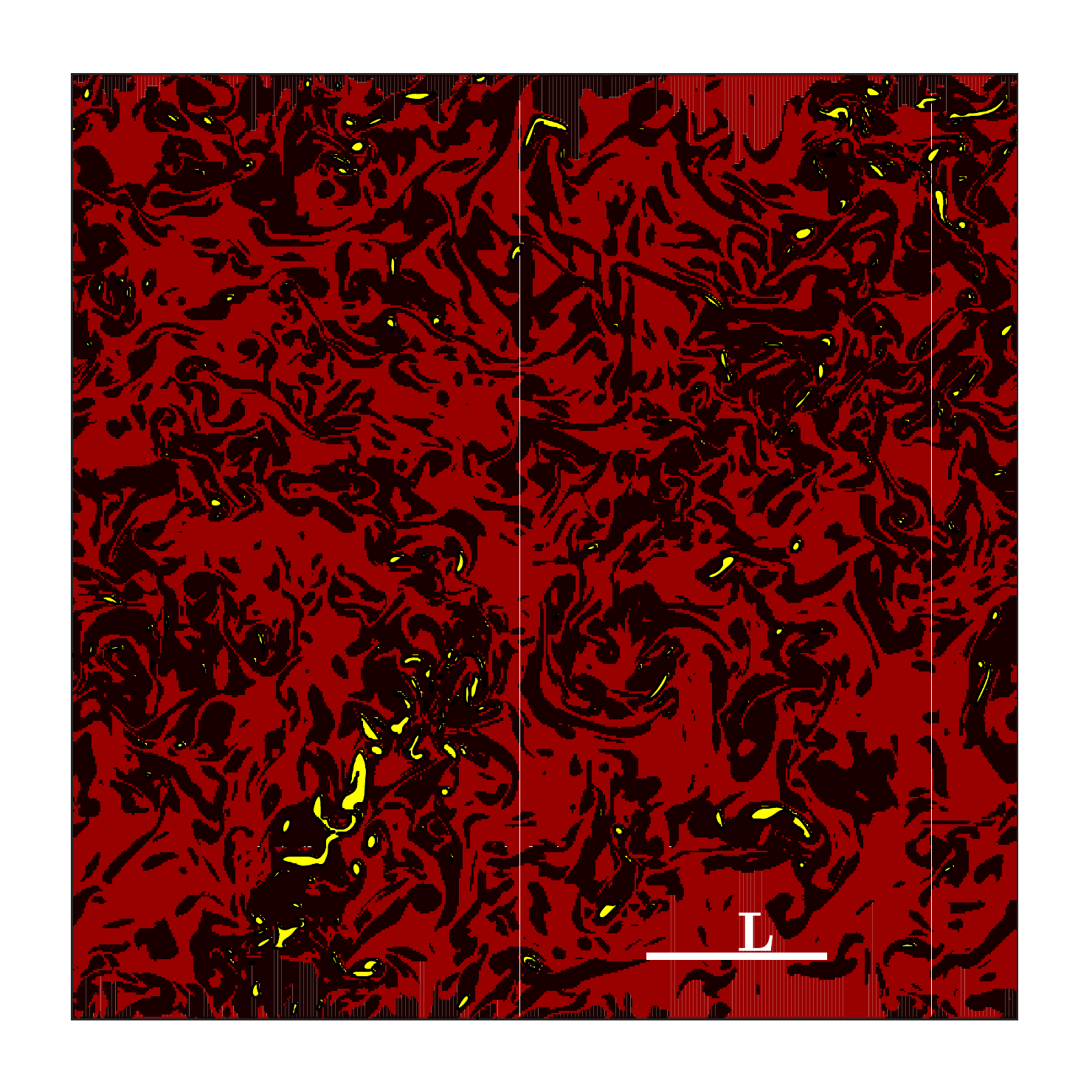}

\caption{(left) Temporal evolution of the average energy dissipation, $\varepsilon(t)$, the surrogate energy dissipation, $\varepsilon_s(t)$, 
and $\Omega^{(3)}(t)$ in a simulation at $Re_{\lambda}=195$. 
Quantities are plotted without their temporal mean and divided by their standard deviation.
(right) Visualisation of the enstrophy field in a plane of the flow for $Re_\lambda=195$. 
The yellow structures correspond to the most intense enstrophy that accounts for $90\%$ of $\langle \Omega^3\rangle$, which 
occupy approximately $1\%$ of the total volume.
The dark read structures correspond to the most intense enstrophy that accounts for $90\%$ of $\langle \Omega\rangle$,
and the light red background to the remaining weak enstrophy that accounts for $10\%$ of $\langle \Omega\rangle$.
The plane shows the full computational domain,
and the white line is equal to $L$.}
\label{fig:1}
\end{figure}

In figure \ref{fig:1}(left), we show the evolution of $\Omega^{(3)}$, of the instantaneous space-averaged dissipation, $\varepsilon$,
and of the instantaneous surrogate dissipation, $\varepsilon_s$, in a time-interval of the simulation at $Re_\lambda=195$.
To compare the three signals, we have subtracted their temporal mean and divided them by their standard deviation.
The dissipation signal fluctuates around its mean in time scales comparable to the integral time scale, 
mirroring the fluctuations of the the surrogate dissipation, which occur earlier. 
The time advancement of the surrogate dissipation with respect to the dissipation
reflects the propagation of large-scale fluctuations down the energy cascade.
This process takes place in scale-local steps of duration consistent with the Kolmogorov theory \citep{cardesa2015temporal},
and is observable in a Lagrangian frame of reference \citep{meneveau1994lagrangian,wan2010dissipation,ballouz2020temporal}.
The characteristic oscillation frequency of the dissipation 
signal is similar in flows with different large-scale forcing \citep{cardesa2015temporal}, 
suggesting that it is a universal signature of inertial dynamics.
An important aspect of figure \ref{fig:1}(left) is that $\Omega^{(3)}$ seems to be advanced with respect to $\varepsilon$, delayed with respect to $\varepsilon_s$, 
and largely correlated to both signals. 

The correlation between $\varepsilon$ and $\Omega^{(3)}$ is remarkable considering that the latter contains
information on the evolution of only a very small fraction of the flow domain, 
which corresponds to events of intensity $\omega^{(3)}\approx70\overline{\langle\Omega\rangle}$ (see table \ref{fig:table}).
A measure of the volume covered by $\Omega^{(p)}$ is given by $\upsilon_{(p)}(\Omega)$,
which we define as the volume fraction occupied by the most intense events of $\Omega$ that account for $90\%$ of
$\langle \Omega^p\rangle$ (similar for $\Sigma$).
In figure \ref{fig:1}(right), we show the isocontours that enclose $\upsilon_{(3)}(\Omega)$
in a flow field at $Re_\lambda=195$. 
These isocontours correspond to the core of the most intense vortices, which
occupy a volume fraction of $\upsilon_{(3)}(\Omega)\approx0.01$ ($1\%$ of the flow domain).
They appear distributed across the domain and separated by distances of the order of $L$.
For comparison, we also show isocontours that enclose $90\%$ of the average enstrophy,
which correspond to structures that spread across the domain 
and occupy a volume fraction of $\upsilon_{(1)}(\Omega)\approx0.5$ ($50\%$ of the domain).
These differences suggest that $\varepsilon$ and $\Omega^{(3)}$ are not directly related.
A reasonable explanation for their correlation comes from $\varepsilon_s$,
which appears to be a precursor of both signals; 
large-scale fluctuations seem to modulate intense events 
in the same way that they modulate the average dissipation.

\begin{figure}
\includegraphics[width=0.6\textwidth]{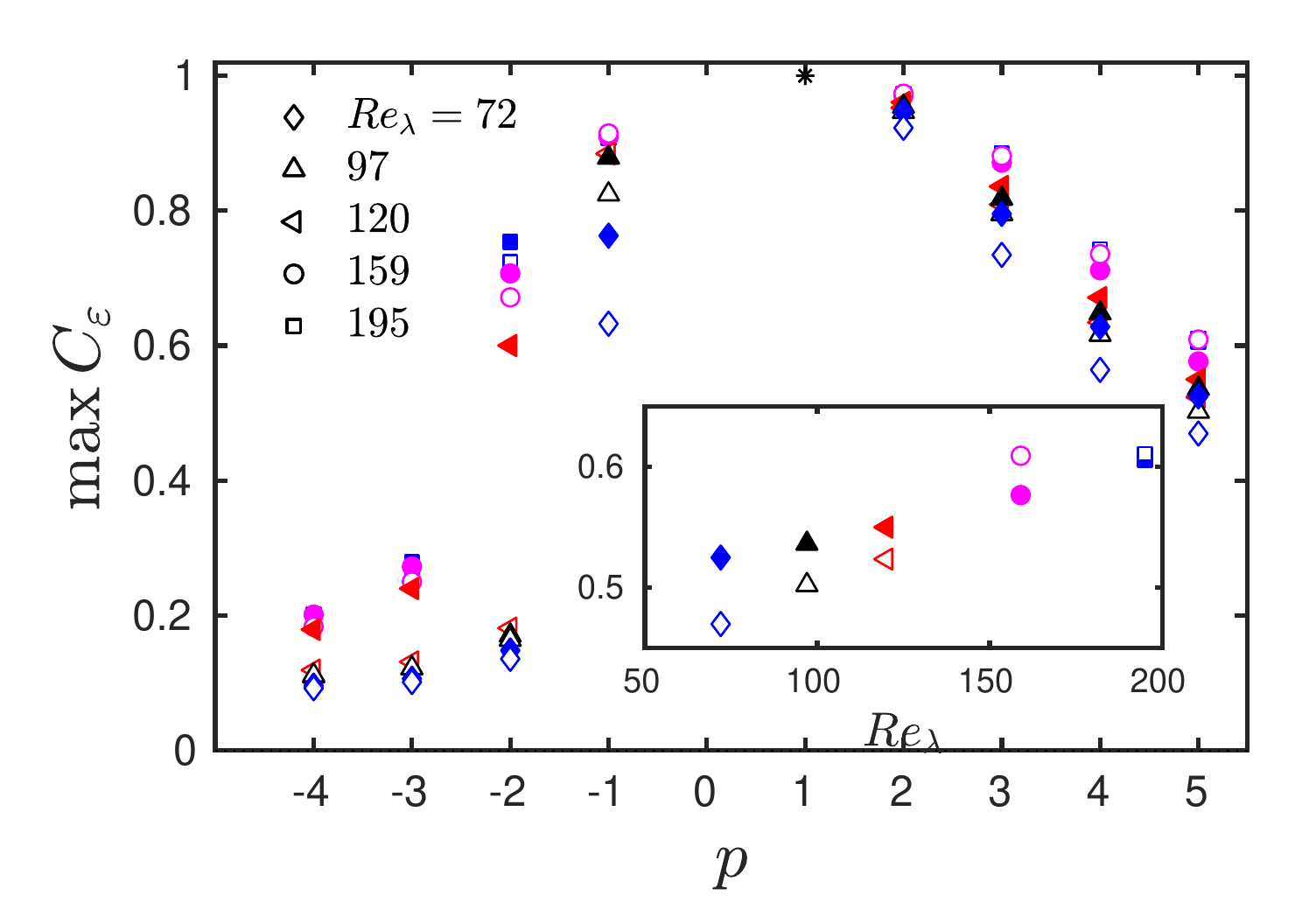}
\caption{Maximum temporal correlation coefficient of $\varepsilon$ with $\Omega^{(p)}$ and $\Sigma^{(p)}$ for different Reynolds numbers.
The inset shows the maximum correlation for $p=5$ as a function of the Reynolds number. Empty symbols correspond to $\Omega^{(p)}$,
and solid symbols to $\Sigma^{(p)}$.
}
\label{fig:2}
\end{figure}

We systematically study this phenomenon and show that it is a persistent signature of the dynamics.
We use the temporal cross-correlation coefficient (TCC) of the dissipation signal, defined as
\begin{equation}
\mathcal C_{\varepsilon}(\tau;\psi)=\frac{1}{T_{s}}\int_0^{T_{s}}\frac{\psi'(t - \tau)\varepsilon'(t)}{\sqrt{\overline{\psi'^2}{\color{white}\cdot}\overline{\varepsilon'^2}}}\mathrm d t,
\end{equation}
where $\psi$ is the test signal, $\tau$ is a time shift, and the prime denotes quantities without temporal average.
In figure \ref{fig:2}, we show the maximum value of $C_{\varepsilon}(\Omega^{(p)})$ and $C_{\varepsilon}(\Sigma^{(p)})$ for different values of $p$ and $Re_\lambda$.
The maxima of the TCC are similar for $\Omega^{(p)}$ and $\Sigma^{(p)}$ and decay for increasing $p$ and $p>1$; 
$\max\mathcal C_\epsilon\approx0.8$ for $p=3$, 
and $\max{C}_\varepsilon\approx0.5$ for $p=5$. 
This last case is significant considering that, for $Re_\lambda=195$,
$\upsilon_{(5)}(\Omega)\approx10^{-4}$, i.e., the events 
represented by $\Omega^{(5)}$ occupy approximately $0.01\%$ of the total flow domain ($\Sigma^{(5)}$ yields comparable results).
In the inset of figure \ref{fig:2},
we show that $\max C_\varepsilon(\Sigma^{(5)})$ increases with $Re_\lambda$,
and that $\max C_\varepsilon(\Omega^{(5)})$ plateaus, 
suggesting that these correlations should persist at higher $Re_\lambda$ 
and that they are not a finite-Reynolds-number effect.
In figure \ref{fig:2}, we have also included the maxima of the TCC for $p<0$
to show that the weak turbulent background is less correlated to the average dissipation than 
the intense events. This indicates that the correlations for $p>0$ are not an statistical artefact of the $p$-means.
For simplicity, we have only 
correlated the $p$-means with $\varepsilon$, but the results are very similar when we consider $\varepsilon_s$.
This is so because $\varepsilon$ and $\varepsilon_s$ are very correlated;
$\max C_\varepsilon(\varepsilon_s)>0.9$ in all cases. 

\begin{figure}
\includegraphics[width=0.45\textwidth]{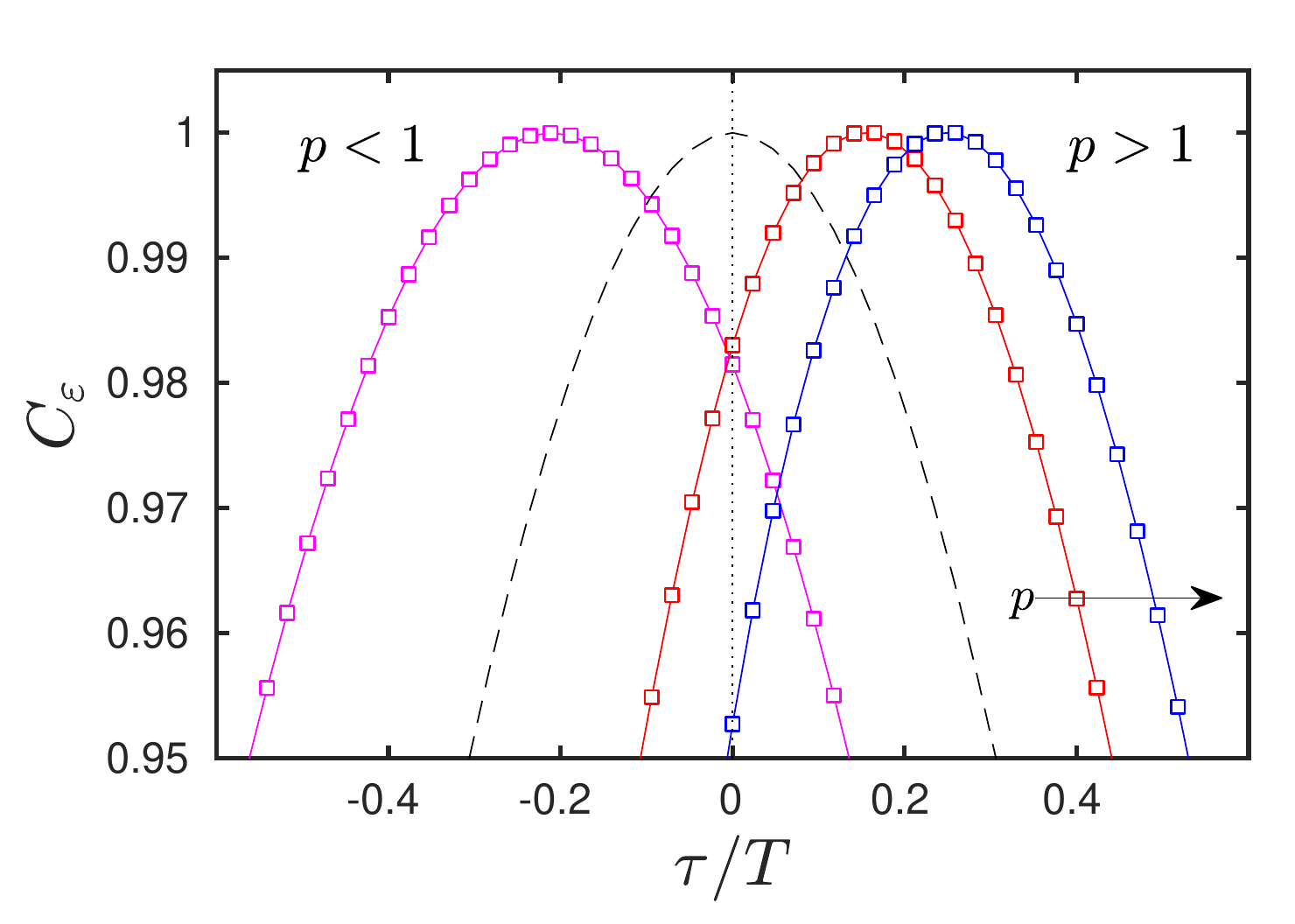}
\includegraphics[width=0.45\textwidth]{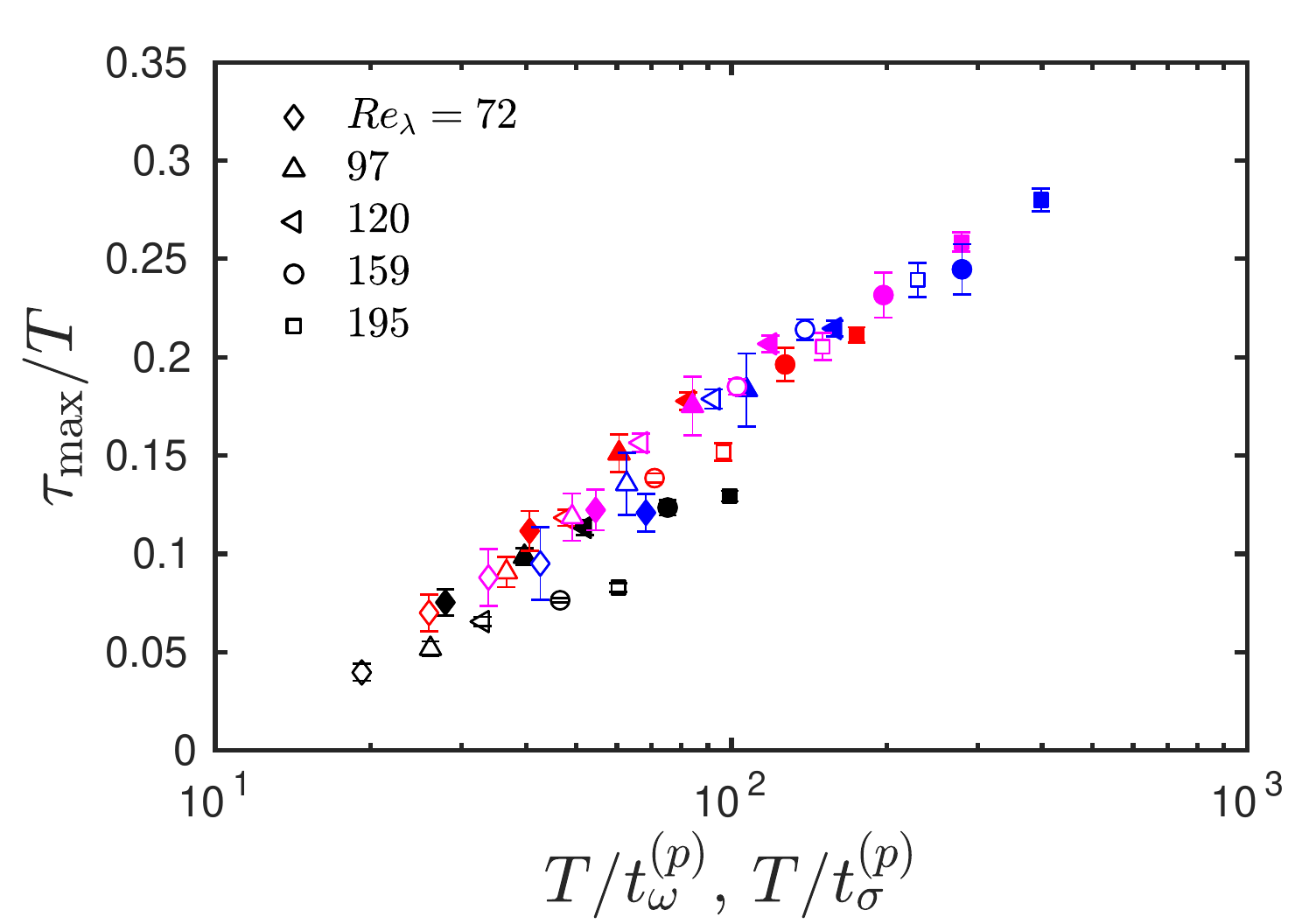}
\caption{
(left) Correlation coefficient of $\varepsilon$ with $\Omega^{(p)}$ for different values of $p$ at $Re_\lambda=195$.
The solid lines with markers correspond to $p=$: (black) -1; (red) 3; (blue) 5. The dashed line corresponds to the 
temporal autocorrelation of $\varepsilon$.
(right) Time shift for which $C_\varepsilon$ is maximum, $\tau_{\max}$, for (solid symbols) $\Omega^{(p)}$ and (empty symbols) $\Sigma^{(p)}$ as a function of the inverse of the characteristic turnover time of each $p$-mean, $t_\omega^{(p)}$ and $t_\sigma^{(p)}$, for different Reynolds numbers.
The colours correspond to $p=$: (black) 2; (red) 3; (magenta) 4; (blue) 5.
The error bars mark the standard deviation of the data obtained by dividing the temporal signal in $4$ subsets.}
\label{fig:3}
\end{figure}

The temporal advancement of $\Omega^{(3)}$ with respect to the dissipation is 
also captured by the TCC.
In figure \ref{fig:3}(left), we show the TCC divided by its maximum as a function of the time shift, $\tau$,
for different values of $p$.
The time shift at which the TCC peaks, $\tau_{\max}$, is positive for the 
$p$-means with $p>1$, meaning that these signals precede on average
the dissipation signal. This temporal advancement grows with increasing $p$.
On the other hand, for $p<1$ the $p$-means are delayed with respect to 
the average dissipation.

We formally calculate $\tau_{\max}$ for different $p$ and Reynolds numbers by fitting each correlation around its maxima with a third-order polynomial.
In figure \ref{fig:3}(bottom), we plot $\tau_{\max}$ against the inverse of the characteristic times of the $p$-means, $t_\sigma^{(p)}$ and $t_\omega^{(p)}$ (see (\ref{eq:22})), and normalise all quantities with the integral eddy-turnover time.
The advancement of $\Omega^{(p)}$ and $\Sigma^{(p)}$ collapses well with this normalisation,
and $\tau_{\max}$ grows as the logarithm on the inverse of $t_\sigma^{(p)}$ and $t_\omega^{(p)}$.
Large values of $\tau_{\max}$ occur for large $p$ and large Reynolds numbers, and reach up to $\tau_{\max}\approx0.3T$.
The advancement of the surrogate dissipation, $\varepsilon_s$, with respect to $\varepsilon$ is approximately $T$ for all Reynolds numbers,
indicating that the $p$-means are delayed with respect to $\varepsilon_s$, 
and that this delay decreases with decreasing $t_\sigma^{(p)}$ and $t_\omega^{(p)}$.

\section{Discussion}

We now summarise and discuss the results presented above.
We have shown that the evolution of intense enstrophy and strain events in the dissipative range of isotropic turbulence, 
as described by the $p$-means, is substantially correlated to large-scale fluctuations and to the average dissipation signal.
Moreover, the fluctuations of the intense events are advanced in time with respect to the dissipation and delayed with respect to large-scale fluctuations,
i.e, intense events of the velocity gradients occur on average before large-scale fluctuations are transformed into dissipation.

These observations admit a causal interpretation.
Correlation does not, in general, imply causation, except in the case of strongly unidirectionally coupled systems \citep{ye2015distinguishing}.
This may be the case in small-scale turbulence.
Recent experiments have shown that intense vorticity synchronises to inertial-range dynamics \citep{vela2021},
suggesting that the relation between the turbulence cascade and intense events is the small scales is
that of a master-slave scenario,
in which intense small-scale fluctuations are driven by inertial-range dynamics.
Probably the most eminent manifestation of this unidirectional coupling is the dissipative anomaly, 
which reflects the control that large-scale dynamics exert on the small scales---through the energy
cascade---to produce the adequate dissipation, even for vanishing viscosity.
In this light, we state that the correlations reported here manifest 
the causal influence of the energy cascade on the intense events in the dissipative range.



This claim is supported on two observations.
First, the temporal correlation between the average dissipation signal (or the surrogate dissipation)
and the most intense structures is hard to explain without the cascade process. 
That a few small-scale structures separated by distances of the order of the integral scale (see figure \ref{fig:1}) 
follow an organised temporal evolution is only conceivable if they emerge from the same large-scale event, 
which reaches the dissipative scales through a cascade process.

Second, the time advancement of the $p$-means with respect to the average dissipation indicates that intense events 
are not caused by small-scale fluctuations, but by events that precede them, 
namely large-scale fluctuations.
%

In line with this idea, we note that the growth of the time advancement with the intensity of the $p$-means, 
and its scaling in integral time units, are consistent with a local multiplicative cascade.
Let us consider an eddy of size $\ell_0$, and assume that it cascades to a scale $\ell_1=\ell_0/2$ in a time proportional to its eddy turnover time,
$t_0=(\ell_0^2/\varepsilon_\ell)^{1/3}$, where $\varepsilon_\ell$ is a measure of the eddy's intensity associated with 
the local energy flux, a quantity that is conserved on average through the cascade 
and translates into a local dissipation when $\ell_n\sim\eta$.
As described by multiplicative cascade models \citep{frisch1978simple}, 
$\varepsilon_\ell$ is unevenly distributed to following generations of eddies,
leading to their multiplicative amplification.
The more amplified an eddy, the faster it cascades to smaller scales, thus advancing weaker eddies. 
This picture is consistent with the results in figure \ref{fig:3}(right).
The scaling of the advancement in integral time units 
suggests that this process does not take place in a single step,
but that it starts in the large scales and accumulates through the cascade.
In agreement with our data, increasing the Reynolds number allows for more cascade steps,
leading to an increased multiplicative amplification of eddies, 
and a faster cascading process of the most amplified ones.
By the same token, eddies that are weaker than the average dissipation
take longer to reach the dissipative scales, as shown in figure \ref{fig:3}(left).

We report in advance that it is possible to qualitatively reproduce our data
using the multifractal cascade model in \citep{meneveau1987simple}, 
and considering the cascade time of eddies as proportional to their turnover time. 
This qualitative agreement suggest that the phenomenological picture
conveyed by multiplicative models is physically sound,
and encourages the pursue of a quantitative agreement 
guided by physically meaningful arguments.
This is a challenging task that requires adding temporal dynamics to the model, 
and it will be addressed elsewhere.


To conclude, we stress a fundamental implication of our results.
We have shown that intense events in small-scale turbulence contain more information on the average evolution of the flow than previously thought \cite{frisch1995turbulence}.
This explains why some models constructed with intense vortices are able to accurately reproduce some global statistics in isotropic turbulence \cite{chorin1988spectrum,she1994universal}, and suggests that, by probing just a few intense structures, it is possible to predict the evolution of the average dissipation field. 
Perhaps this property also extends to other statistics of interest, which could be exploited for predictive applications or modelling.   
From an inverse perspective, our findings could also be leveraged to develop strategies to predict and control extreme events 
in turbulence, a problem of applied interest which has proven very challenging \citep{sapsis2020statistics}.

\bibliography{references}

\end{document}